\documentclass[%
 reprint,
superscriptaddress,
nofootinbib,
 amsmath,amssymb,
 aps,
 prl,
floatfix,
]{revtex4-2}

\usepackage{enumitem}

\usepackage{hyperref}
\usepackage{graphicx}
\usepackage{dcolumn}
\usepackage{bm}

\usepackage{epsfig}
\usepackage{amsmath}
\input{epsf}
\usepackage{psfrag}
\usepackage{xcolor}
\usepackage{pdfpages}
\usepackage[normalem]{ulem}
\usepackage{subcaption} 

\makeatletter
\AtBeginDocument{\let\LS@rot\@undefined}
\makeatother
\newcommand{\bea}{\begin{eqnarray}}
\newcommand{\eea}{\end{eqnarray}}

\newcommand{\nn}{\nonumber}

\begin{document}

\title{
Universality in the Near-Side Energy-Energy Correlator
}
 
\author{Xiaohui Liu}
 \email{xiliu@bnu.edu.cn}
 \affiliation{Center of Advanced Quantum Studies, School of Physics and Astronomy, Beijing Normal University, Beijing, 100875, China}
 \affiliation{Key Laboratory of Multi-scale Spin Physics, Ministry of Education, Beijing Normal University, Beijing 100875, China}

\author{Werner Vogelsang}%
 \email{werner.vogelsang@uni-tuebingen.de}
\affiliation{Institute for Theoretical Physics,
                Universit\"{a}t T\"{u}bingen,
                Auf der Morgenstelle 14,
                D-72076 T\"{u}bingen, Germany}
 
\author{Feng Yuan}%
 \email{fyuan@lbl.gov}
\affiliation{Nuclear Science Division, Lawrence Berkeley National Laboratory, Berkeley, CA 94720, USA}%
\affiliation{Institute for Theoretical Physics,
                Universit\"{a}t T\"{u}bingen,
                Auf der Morgenstelle 14,
                D-72076 T\"{u}bingen, Germany}

\author{Hua Xing Zhu}%
 \email{zhuhx@pku.edu.cn}
\affiliation{School of Physics, Peking University, Beijing 100871, China}%
\affiliation{Center for High Energy Physics, Peking University, Beijing 100871, China}

\begin{abstract}
We investigate the energy-energy correlator (EEC) of hadrons produced on the same side in 
$e^+e^-$ annihilation or in leading jets in $pp$ collisions. We observe a remarkable universality of
the correlator. Using a non-perturbative transverse momentum dependent (TMD) fragmentation function 
to model the transition from the ``free-hadron" region to the perturbative collinear region, 
we are able to describe the near-side shapes and peaks over a wide range
of energy for both the $e^+e^-$ annihilation and the $pp$ jet substructure measurements in terms of just two parameters. We present further predictions for the ratio of the projected three-point energy correlator to the EEC. 
The excellent agreement between our calculations and the experimental data may provide new insights into the role of non-perturbative physics for EECs, and suggests the possibility of exploring non-perturbative TMDs using theoretical tools developed for the energy correlators.
\end{abstract}

\maketitle

\textbf{\textit{  Introduction.}} Since their introduction more than four decades ago~\cite{Basham:1978bw}, hadronic energy-energy correlators (EECs) measured in 
various high energy scattering processes have become important precision observables to study strong interaction physics and to investigate the
ermergence of hadronic final states in QCD~\cite{Collins:1981uk,Ali:1982ub,Clay:1995sd,deFlorian:2004mp,DelDuca:2016csb,Tulipant:2017ybb,Kardos:2018kqj,Moult:2018jzp,Dixon:2018qgp,Dixon:2019uzg}. In recent years, there has been a renaissance of interest in EECs due to their applications in QCD jet substructure~\cite{Chen:2020vvp,Chen:2020adz,Komiske:2022enw,Lee:2022ige,Chen:2023zlx,Chen:2019bpb,Chicherin:2024ifn,He:2024hbb}, top quark physics~\cite{Holguin:2023bjf,Holguin:2024tkz,Xiao:2024rol}, jet formation in heavy ion collisions~\cite{Andres:2022ovj,Andres:2023xwr,Andres:2023ymw,Yang:2023dwc,Andres:2024ksi,Barata:2023bhh,Barata:2023zqg,Bossi:2024qho,Singh:2024vwb}, connection to TMD physics~\cite{Gao:2019ojf,Li:2020bub,Ebert:2020sfi,Li:2021txc,Kang:2023big,Gao:2024dbv,Kang:2024otf}, and heavy flavored hadron fragmentation~\cite{Craft:2022kdo,Chen:2024nfl}. This renewed focus has led to a suite of new measurements across different colliding beams, from proton-proton collisions~\cite{CMS:2024mlf,ALICE:2024dfl} to relativistic heavy ion collisions~\cite{Tamis:2023guc,CMS:2024ovv}. Beyond their phenomenological significance, EECs, as correlation functions of energy flow operators (also known as ANEC operators), have important applications in diverse fields such as conformal field theory~\cite{Hofman:2008ar,Belitsky:2013xxa,Kravchuk:2018htv,Kologlu:2019mfz}, quantum gravity~\cite{Gao:2000ga}, quantum information~\cite{Faulkner:2016mzt}, and quantum chaos~\cite{Hartman:2016lgu}. As a result of the recent developments, EECs have emerged as a powerful interdisciplinary tool, forging connections between formal theory, particle and nuclear phenomenology, and collider experiments.

Taking the energy-energy correlator in the $e^+e^-$ annihilation process as an example, we define
\begin{eqnarray}
{\rm EEC}(\chi) 
&\equiv&    \frac{1}{\sigma}\sum_{a,b}\int \frac{E_aE_b}{Q^2} d\sigma(e^+e^-\to h_ah_bX)\nn\\
&&\times  
\delta\left(\chi - \theta_{ab}\right)
\ ,\label{eq:eecdef}
\end{eqnarray}
where $\sigma$ is the total cross section and $Q=\sqrt{s}$ the c.m.s. collision energy. The sum runs over all pairs of
hadrons $h_a,h_b$ with associated energies $E_a,E_b$ and with 
$\cos\theta_{ab}=\boldsymbol{p}_a \cdot \boldsymbol{p}_b/|\boldsymbol{p}_a||\boldsymbol{p}_b|$. 
Two regions in $\chi$ are of particular interest: the {\it near side} with $\chi\to 0$, and the {\it away side} (or, back-to-back region), for which 
$\pi-\chi\to 0$. 

On the away side, when $\chi\to\pi$ the EEC is dominated by 
soft and collinear gluon radiation. Consequently, an appropriate TMD factorization has been developed and applied~\cite{Collins:1981uk,Collins:1981uw,Collins:1981va,Collins:1981zc} 
which entails an all-order resummation of double logarithms in $1+\cos(\chi)$~\cite{Collins:1984kg,deFlorian:2004mp,Moult:2018jzp}. This TMD resummation has been included in the phenomenological studies of the EEC measurements in $e^+e^-$ annihilation~\cite{deFlorian:2004mp,Tulipant:2017ybb,Kardos:2018kqj,Ebert:2020sfi}.  
In the perturbative near-side regime, the EEC is dominated by collinear radiation, and a resummation of large logarithms can be further included~\cite{Konishi:1979cb,Dixon:2019uzg} with additional contributions from power corrections and/or non-perturbative hadronization effects~\cite{Korchemsky:1994is,Korchemsky:1999kt,Belitsky:2001ij,Abbate:2010xh,Schindler:2023cww,Jaarsma:2023ell,Lee:2024esz,Chen:2024nyc}. A striking feature is that
toward very small angle the EEC has been shown to capture the transition from asymptotically free partons to confined hadrons~\cite{Komiske:2022enw}. Furthermore, beyond this transition point, the EEC exhibits a behavior expected for ``free hadrons'',namely a flat scaling behavior with vanishing anomalous dimension.
Theoretical arguments from strong coupling limits of 
formal field theory computations have also been applied in this region~\cite{Hofman:2008ar,Hatta:2008tx,Hatta:2008st,Chicherin:2023gxt,Chen:2024iuv,Csaki:2024joe}. However, in the literature there are only a few model suggestions to describe the near-side EEC, such as a parameterization based on the ``free hadrons'' assumption~\cite{Andres:2024ksi}, or applying a kinematic cutoff in the perturbative calculations~\cite{Yang:2023dwc}. 

We note that the near-side region has also been addressed in the literature in terms
a (collinear) di-hadron fragmentation picture~\cite{deFlorian:2003cg}. Further developments of this have mainly 
focused on resonance kinematics and aimed at using double hadron production as a novel probe of hadronic structure in various collision experiments~\cite{Jaffe:1997hf,Bianconi:1999cd,Bacchetta:2002ux,Bacchetta:2012ty,Zhou:2011ba,Cocuzza:2023vqs,Pitonyak:2023gjx}. 
However, the application of this picture to the EEC is not straightforward and has not been discussed so far.

In the present paper, we propose a different approach. Inspired by previous discussions of non-perturbative contributions to event shape observables and EECs~\cite{Korchemsky:1994is,Korchemsky:1999kt,Belitsky:2001ij,Schindler:2023cww}, we will develop an effective model
framework to compute the near-side EEC from a {\it universal} non-perturbative TMD fragmentation contribution. As we will show, this relatively simple approach is remarkably successful in phenomenology. Description of the near-side EEC by a TMD appears quite natural: When $\chi Q \lesssim \Lambda_{\rm QCD}$, the correlations of hadrons must become sensitive to the relative internal transverse motion of the parton progenitors,
which is akin to what is typically captured by a TMD. We will discuss in how far this simple picture can be made more rigorous. 

It is immediately clear that such a near-side TMD fragmentation function must be fundamentally different from those for the away side. This is because
there are no perturbative Sudakov effects on the near side since soft gluon radiation is suppressed in the energy-weight measurements. Therefore, for the near-side EEC 
we expect a {\it non-perturbative} TMD fragmentation function framework to be appropriate. We find that it includes 
a scale dependent contribution, which plays a critical role in the description of the near-side EEC from $e^+e^-$ annihilation and jet-substructure in proton-proton collisions across energy scales of different orders of magnitude. Most importantly, we argue that this scale term is closely related to the Collins-Soper evolution kernel for the TMD resummation. The latter has attracted great attention from phenomenology~\cite{Scimemi:2017etj,BermudezMartinez:2022ctj,Boussarie:2023izj} and lattice calculations in recent years~\cite{Shanahan:2020zxr,Shanahan:2021tst,Avkhadiev:2023poz,Avkhadiev:2024mgd,LatticePartonLPC:2022eev,LatticePartonLPC:2023pdv,Schlemmer:2021aij,Shu:2023cot}.

\begin{figure}[htbp]
  \begin{center}
   \includegraphics[scale=0.34]{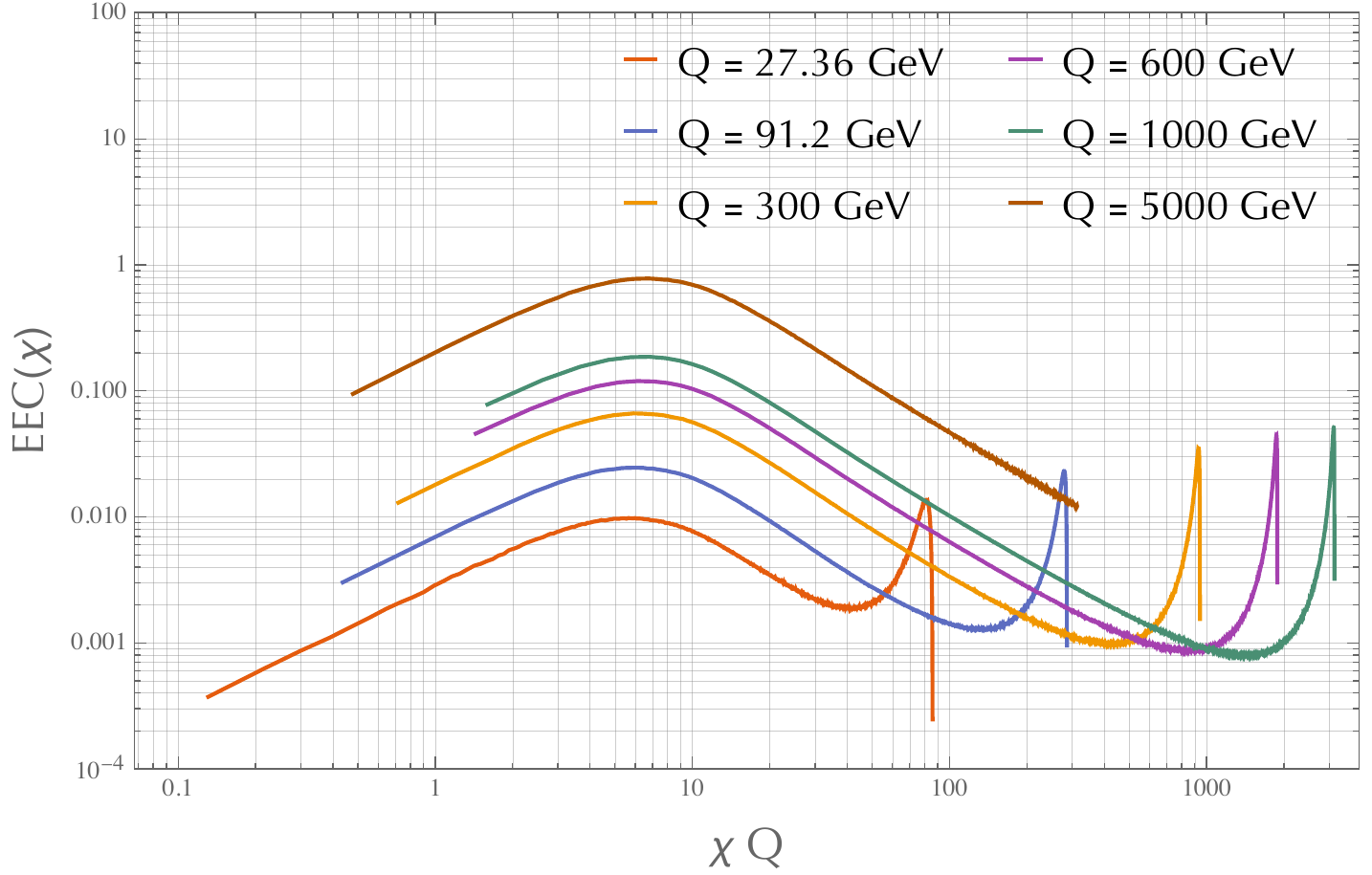} 
\caption{EEC spectrum in $e^+e^-$ annihilation for different values of $Q$ as a function of $\chi Q$. }
  \label{fg:zoom-small-theta}
 \end{center}
  \vspace{-5.ex}
\end{figure}

\textbf{\textit{Non-perturbative features of EEC when $\chi\to 0$.}} 
In Fig.~\ref{fg:zoom-small-theta}, we plot the EEC in $e^+e^-$ annihilation defined in Eq.~(\ref{eq:eecdef}) as a function of $\chi Q$, focusing on the limit $\chi \to 0$. Here we are using a Pythia 8.245 simulation~\cite{Sjostrand:2014zea}, which as will be shown below is consistent with experimental data. As one can readily see, 
the peaks at small $\chi$ show a much more pronounced rise with increasing $Q$ than those at $\chi \to \pi$. This relative suppression in the $\chi \to \pi$ region is well understood as a consequence of the perturbative Sudakov factor relevant there. By contrast, the rapidly increasing peaks at small angles strongly support the absence of the perturbative Sudakov factor as $\chi \to 0$, corroborated by both the factorization formalism~\cite{Dixon:2019uzg} and explicit calculation~\cite{Dixon:2018qgp}.     

Figure~\ref{fg:zoom-small-theta} also clearly shows the transition from the ``parton phase" to the ``free-hadron phase" as $\chi$ becomes small, a feature discussed extensively in 
the previous literature~\cite{Komiske:2022enw,CMS:2024mlf, ALICE:2024dfl}. In the ``free hadron" domain, the EEC spectrum scales as ${\rm EEC}(\chi) \propto \chi$, indicating a uniform distribution of energy flow over the solid angle $2\pi d\cos\chi $ as $\chi \to 0$. Additionally, the transition point (the position of the small-angle peak) remains almost independent of $Q\chi$ over several orders of magnitude in $Q$. It is important to note that the transition cannot be described by perturbative calculations which predict a continuously growing spectrum as $\chi \to 0$~\cite{Dixon:2019uzg}. Rather, it must be explained by a non-perturbative mechanism. 

Further features of the EEC are discussed in the Supplemental Material to this letter.

\textbf{\textit{TMD model description of EEC when $\chi \to 0$.}}
To establish a connection to the non-perturbative TMD fragmentation studied in the literature, we consider an arbitrary hadron $h$ and the energy flow in the non-perturbative region $\chi \to 0$ surrounding the hadron. In this regime, we may assume that the transverse momentum $p_T$ associated with any energy flow with respect to this hadron is dominantly driven by uncorrelated non-perturbative soft radiation. Writing the corresponding normalized differential cross section in terms of an effective non-perturbative TMD fragmentation function, we then have
\bea\label{eq:NPff} 
\frac{1}{\sigma_h}
\frac{d^3\sigma}{dz d^2{\bm p}_T}& \equiv& D_h(z,p_T) 
=  d_h(z) 
\int \frac{d^2{\bm b}}{(2\pi)^2}
e^{-i{\bm b} \cdot {\bm p}_T} \nonumber \\
&& \hspace{0.ex} 
\times 
\sum_n \frac{1}{n!}
\prod_{i=1}^n \int [d k_i] M^{\rm NP}(k_i) \left(e^{i{\bm b}\cdot {\bm k}_{ti}} -1 \right)
\nonumber \\ 
&=&d_h(z)
\int \frac{db}{2\pi} b J_0(p_T b)  e^{-S_{NP}(b,\mu)}\,.
\eea 
In these expressions, $d_h(z)$ is a function of the momentum fraction $z$ carried by hadron $h$, and $M^{\rm NP}$ is the matrix element for emission of non-perturbative soft quanta. 
In the last equality, we have introduced the non-perturbative part of the Collins-Soper kernel,
\bea\label{eq:collins} 
S_{\rm NP}(b,\mu) = - 
\int [dk]  
M^{\rm NP}(k) 
\left(e^{i{\bm b}\cdot {\bm k}_{t}} -1  
\right)  \,. 
\eea 
We note that since the phase space integral leads to $\int [dk]  
\propto \int^\mu \frac{dk^+}{k^+}  
\propto \ln \mu 
$, the Collins-Soper kernel $S_{\rm NP}$ depends on a scale $\mu$. This will become important for explaining the 
rapid rise of the peaks in the small $\chi$ limit. 

Summing over all hadrons propagating in the same direction we can now effectively model the EEC for small $\chi$ by the TMD distribution: 
\begin{equation}\label{eq:model}
{\rm EEC}(\chi) |_{\chi\to 0}= (2\pi)E_J^2  \chi
\sum_h \int dz \, 
c^3   
 z D_h(z,cE_J\chi )\,,
\end{equation}
where we have used $p_T \approx c E_J \chi$ with $E_J=Q/2$. The parameter $c$ characterizes the fraction of $E_J$ that
populates the small-angle region around $h$. As the behavior of the EEC in Fig.~\ref{fg:zoom-small-theta} 
indicates, for small $\chi$ the energy flow is nearly uniformly distributed. We hence expect $c$ to be approximately constant.   
Its value may depend on the jet radius $R$; however, we assume $R$ to be sufficiently large for this dependence to be negligible. 
Also, for simplicity, we have neglected a possible scale dependence of $c$. 

We next follow~\cite{Sun:2014dqm} to parameterize the non-perturbative TMD as
\bea 
S_{NP}(b,\mu) = 
\frac{g_1}{z^2} b^2
+ 
\frac{g_2}{2}\ln\left(\frac{b}{b_\ast}\right)
\ln\frac{\mu}{\mu_0} \,,
\eea 
where $b_\ast = b/\sqrt{1+b^2/b_{\rm max}^2}$. 
In the above equation, the 
term with the $g_2$ parameter represents the energy dependence in our results, following the evolution of the Collins-Soper kernel. We will see that this term plays a crucial role in predicting the correct behavior of the near-side EEC as a function of energy. We set $\mu = cE_J$. Another plausible choice is to set $\mu = R E_J$ following the philosophy of factorization theorem~\cite{Lee:2022ige}. While these two options will not lead to visible differences within the kinematic region we will study, we stick to $\mu = cE_J$ in the rest of the work. Inserting into Eq.~(\ref{eq:model})
we find
\bea 
{\rm EEC(\chi)}|_{\chi \to 0}
&=& c^3 N E_J^2 \chi \int db \, 
b\, J_0(c E_J \chi b) e^{-\frac{g_1}{c^2}b^2}\nn\\
&&\times e^{- \frac{g_2}{2}\ln\left(\frac{b}{b_\ast}\right)
\ln\frac{cE_J}{\mu_0}  } \,,
\eea 
where the normalization $N e^{\frac{g_1}{c^2} b^2} \approx \int dz \sum_h z d_h(z) e^{\frac{g_1}{z^2} b^2}$ remains to be determined. 
This completes our model considerations and is our main result, which we will now apply to the EEC data and simulations. We note that the slope of 
the EEC in our model for $\chi\to 0$ is related to the zeroth moment of the Collins-Soper kernel:
\bea 
m_0 = c^3NE_J^2 \int db b  e^{-\frac{g_1}{c^2}b^2}
e^{- \frac{g_2}{2}\ln\left(\frac{b}{b_\ast}\right)
\ln\frac{cE_J}{\mu_0}  } \,. 
\eea

\textbf{\textit{Fixing model parameters and comparison to experimental data.}} 
In principle, the model parameters could be obtained from a global fit to experimental data. Fortunately, most parameters are related to the quark TMD Collins-Soper kernel and have already been determined using semi-inclusive deep-inelastic scattering (SIDIS) data~\cite{Sun:2014dqm}. For the quark EEC, we follow Ref.~\cite{Sun:2014dqm} to set $g_1 = 0.042{\rm GeV}^2$, $g_2 = 0.84$, $\mu_0 = 1.55 \>{\rm GeV}$, and $b_{\rm max} = 1.5\> {\rm GeV}^{-1}$. This leaves only two free parameters, the normalization $N$, and the energy fraction $c$, which we determine by reproducing the quark-jet EEC distribution at $300\>{\rm GeV}$, resulting in $N = 0.8670$ and $c = 0.346$. The parameters for the gluon distribution are less constrained, for simplicity we just rescale $g_2 \to \frac{C_A}{C_F} g_2$ for gluons, keeping all other parameters identical to the quark ones. 

\begin{figure}[h]
  \begin{center}
   \includegraphics[scale=0.34]{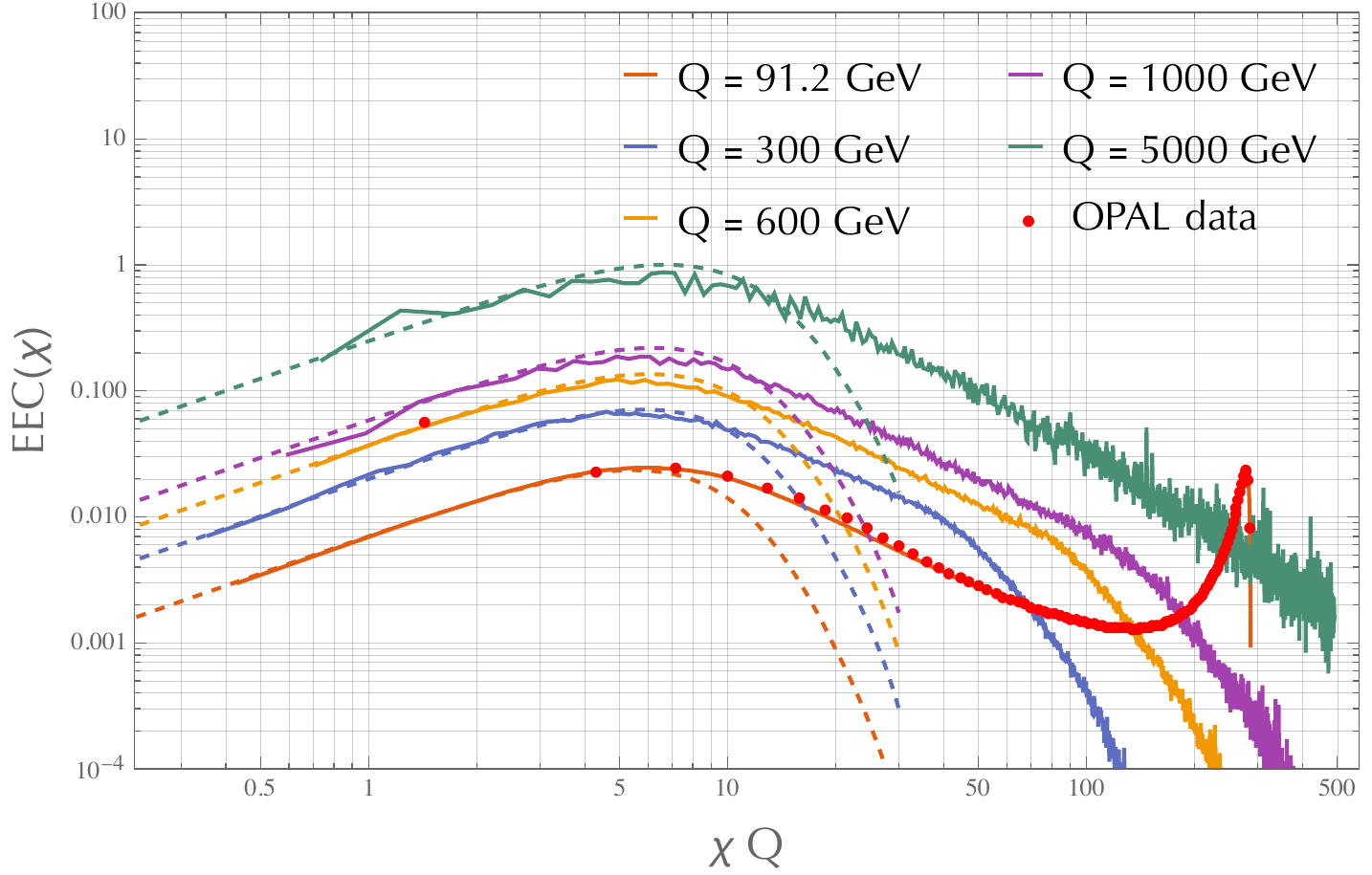} 
\caption{Comparison between the quark EEC obtained by Pythia simulation (solid lines) or OPAL data \cite{OPAL:1993pnw} (red dots) with our model prediction (dashed curves).}
  \label{fg:model-q1}
 \end{center}
  \vspace{-5.ex}
\end{figure}

We validate the model by comparing it to Pythia simulations~\cite{Sjostrand:2014zea} and experimental data. The comparison to Pythia for the EEC constructed from quark-initiated jets is shown in 
Fig.~\ref{fg:model-q1}, where the jet energy $E_J \in (0.98\frac{Q}{2},1.02\frac{Q}{2})$. Details on the Pythia simulation are given in the Supplemental Material. We show results for various $Q$, including $Q = 91.2\>{\rm GeV}$ where we also compare to OPAL experimental data \cite{OPAL:1993pnw}. The model predictions are shown as dashed curves, with variations in $Q$ being the only factor influencing these predictions, as all parameters have been previously determined. Figure~\ref{fg:model-q1} indicates good agreement between our model and the Pythia/OPAL results in the non-perturbative regime. Notably, the model captures both the $Q$ dependence and the position of the peak in the transition region. Beyond the peak, outside the ``free hadron" regime, the model naturally fails to describe the data. In this region, a dihadron fragmentation picture and/or matching to a perturbative calculation will be required. We note that we have also compared our model to Pythia simulations for the EEC distribution measured in gluon jets, finding a similarly good agreement. Details are given in the Supplemental Material.

\begin{figure*}[t]
  \begin{center}
   \includegraphics[scale=0.4]{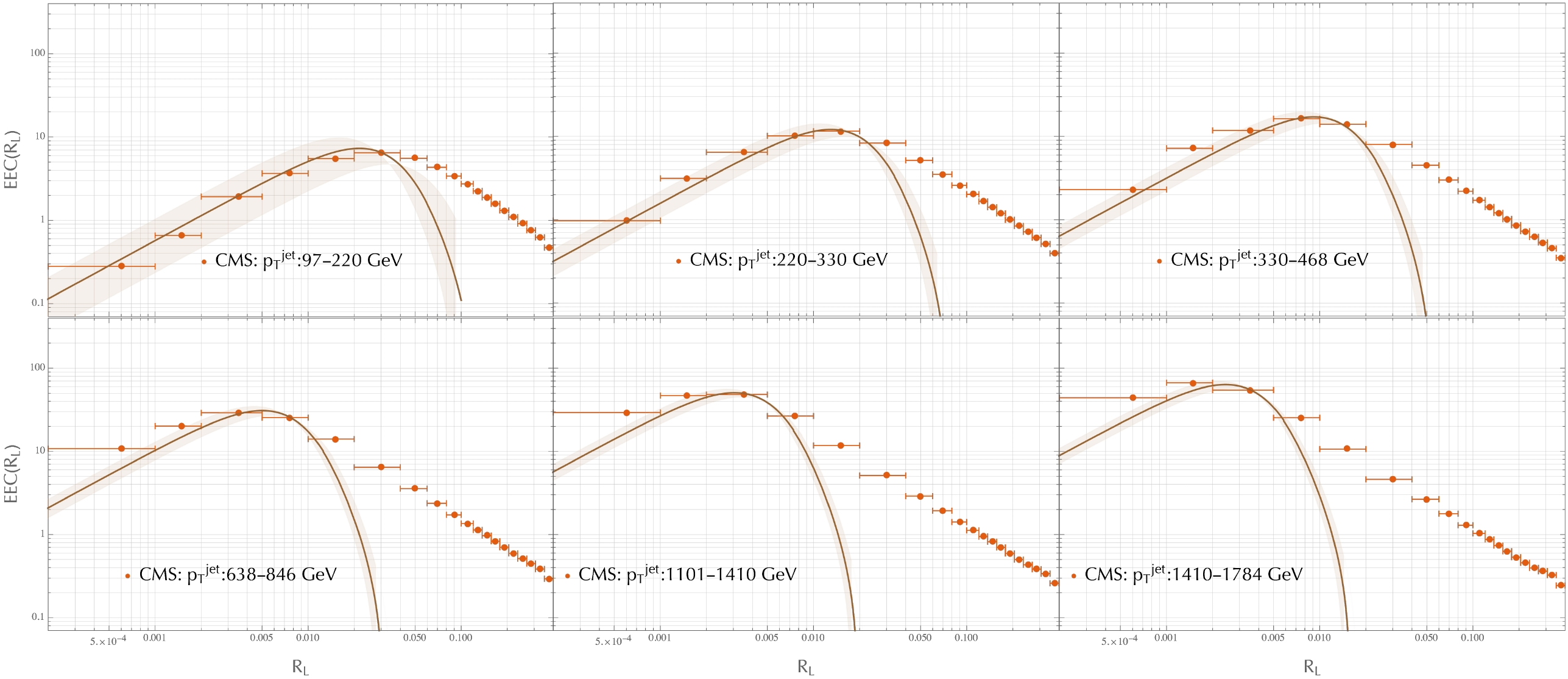} 
\caption{Comparison of our TMD model predictions (solid curves) to CMS data \cite{CMS:2024mlf} at $\sqrt{s}=13$ TeV (red dots). Here we follow the CMS notation where $R_L = \chi$. }
  \label{fg:cms-eec1}
 \end{center}
  \vspace{-5.ex}
\end{figure*}

We next turn to $pp$ collisions and test our model predictions against recent 
CMS measurements~\cite{CMS:2024mlf} at $\sqrt{s}=13$ TeV.
Figure~\ref{fg:cms-eec1} shows the EEC measured in events with at least two jets, defined via the anti-$k_T$ algorithm with $R=0.4$, following CMS to require pseudorapidity $|\eta_J| < 2.1 $ for each jet and an azimuthal separation of $\Delta|\phi| > 2.0$ between the jets. We estimate the relative quark and gluon jet fractions using a Pythia simulation as shown in Table \ref{table:jet_fractions}.
Our model prediction is then given by $\Sigma = f_g \Sigma_{g-{\rm jet}} + f_q \Sigma_{q-{\rm jet}}$ with the scale $E_J$ approximated by the medium value of $p_T$ in each $p_{{\rm jet},T}$ bin. The results are represented by the brown solid curves in Fig.~\ref{fg:cms-eec1}, where the band is derived by adjusting $E_J$ to the upper and lower boundaries of each $p_{{\rm jet},T}$ bin. The close match between the model predictions and the CMS data clearly suggests the universality of
the mechanism for EEC in high-energy collisions. We note that we have also compared to recent ALICE data~\cite{ALICE:2024dfl} and found a similar agreement for the shape in the small-$\chi$ region, see the supplementary material.

\begin{table}[h]
\centering
\begin{tabular}{|c|c|c|}
\hline
$p_{\text{jet},T}$ range (GeV) & $f_g$ & $f_q$ \\
\hline
97 -- 220 & 0.82 & 0.18 \\
220 -- 330 & 0.80 & 0.20 \\
330 -- 468 & 0.78 & 0.22 \\
638 -- 846 & 0.71 & 0.29 \\
1101 -- 1410 & 0.69 & 0.31 \\
1410 -- 1784 & 0.64 & 0.36 \\
\hline
\end{tabular}

\caption{Relative quark ($f_q$) and gluon ($f_g$) jet fractions for different $p_{\text{jet},T}$ ranges.}
\label{table:jet_fractions}
\end{table}

\begin{figure}[htbp]
  \begin{center}
   \includegraphics[scale=0.32]{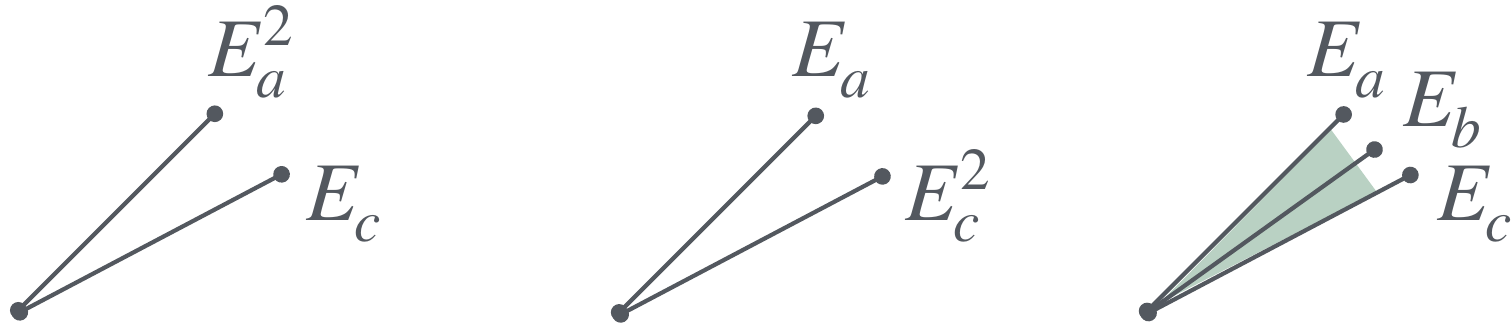} 
\caption{Configurations that contribute to E3C. The three-point configuration is proportional to the area (shaded area) set by $\chi$.}
  \label{fg:e3c}
 \end{center}
  \vspace{-5.ex}
\end{figure}

\begin{figure}[htbp]
  \begin{center}
   \includegraphics[scale=0.43]{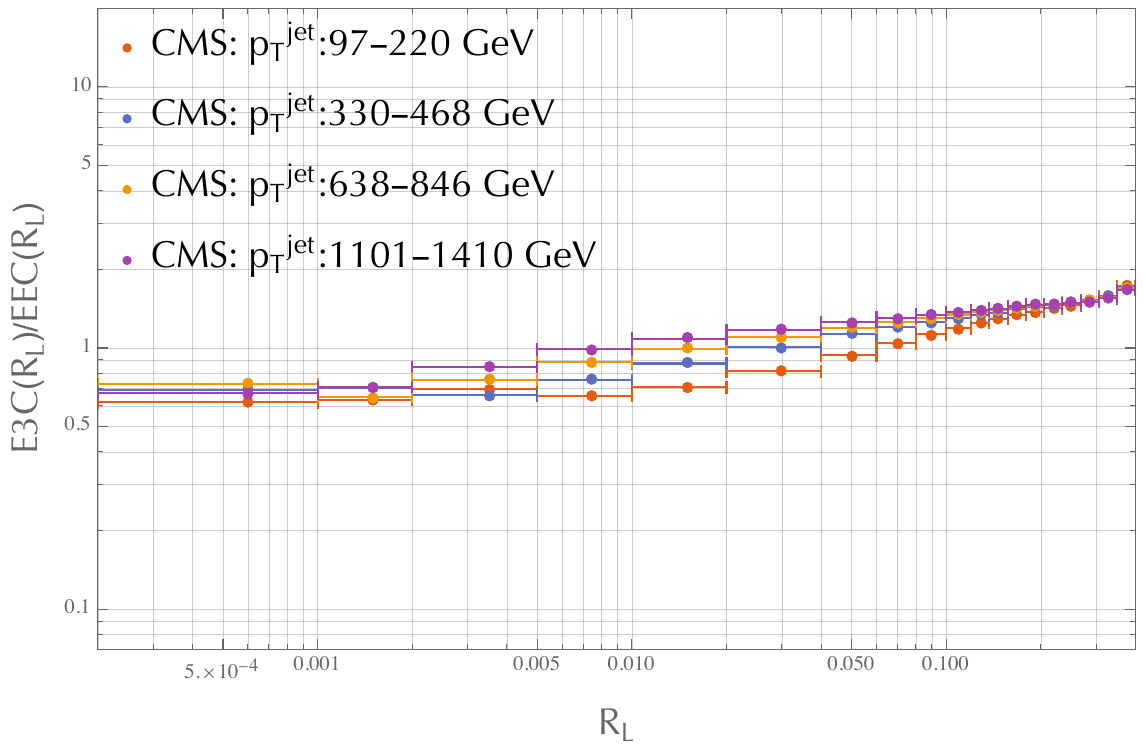} 
\caption{Ratio E3C/EEC calculated from CMS data~\cite{CMS:2024mlf}, where $R_L = \chi$.}
  \label{fg:e3c_eec_cms}
 \end{center}
  \vspace{-5.ex}
\end{figure}

A further prediction of the model may be obtained for the projected E3C distribution~\cite{Chen:2020vvp}, defined as 
\bea 
{\rm E3C}(\chi) 
\equiv \sum_{a,b,c}\int \frac{E_aE_bE_c}{\sigma E_J^3} d\sigma 
\delta \big(\chi - \max(\theta_{ab},\theta_{ac},\theta_{bc})\big) \,.\quad
\eea 
For $\chi\neq 0$ this three-point correlator receives contributions from two-point configurations (e.g., $a = b \ne c$) via the term  
$\frac{E_a^2 E_c}{E_J^3} \delta(\chi - \theta_{ac})$, shown by the two left diagrams in Fig.~\ref{fg:e3c}. Meanwhile, 
the genuine three-point configuration (third diagram in Fig.~\ref{fg:e3c}) is proportional to the area $A \sim  \chi^2$, so that this contribution quickly vanishes as $\chi \to 0$ in the TMD model, leading to the prediction
\bea 
\frac{{\rm E3C}(\chi) }{{\rm EEC}(\chi)} |_{\chi \to 0}
= 2c  \approx 0.7 \,. 
\eea 
This behavior was observed in recent CMS~\cite{CMS:2024mlf} (see Fig.~\ref{fg:e3c_eec_cms}) and ALICE measurements~\cite{ALICE:ALI-PREL-557422}. 
It is remarkable that the value $2c \approx 0.7$ agrees well with the data even though it was derived from an apparently unrelated physical quantity. 
In the Supplemental Material, we also provide results for the ratio from Pythia simulation,
as well as predictions of our model for $\chi\neq 0$.

\textbf{\textit{Conclusions.}} 
We have presented a model for the non-perturbative behavior of the energy-energy correlator in the small $\chi $ limit, linking the small-angle correlation to the well-known Collins-Soper kernel and to non-perturbative TMD fragmentation functions. With most parameters fixed by established results in the literature on SIDIS, only 
two free parameters remain that can be determined simultaneously by fitting the EEC distribution in $e^+e^-$ annihilation at a specific energy. Remarkably, this approach 
describes the non-perturbative small-angle regime of the EEC well over several orders of magnitude in energy. 
We have then compared the model to LHC data for the EEC in leading-jet production in $pp$ collisions, observing similarly good
agreement and demonstrating the universality of the EEC. 
Additionally, our model successfully predicts the E3C to EEC ratio as $\chi \to 0$, compared to both Pythia simulation~\cite{Sjostrand:2014zea} and to CMS~\cite{CMS:2024mlf} and ALICE data~\cite{ALICE:2024dfl,ALICE:ALI-PREL-557422}.   

The results of our paper illustrate the interplay between the EEC and non-perturbative physics associated with hadronization in collider experiments. 
Specifically, our model suggests a close connection between the non-perturbative behavior of the EEC in the near-side and hadronic intrinsic TMD structures. This in turn may
indicate the exciting possibility of understanding non-perturbative TMDs using formal field theoretical tools in the strong coupling limits~\cite{Hofman:2008ar, Hatta:2008tx,Chen:2024iuv,Csaki:2024joe}. We expect our findings to stimulate further studies on the transition regions in various collision systems such as $pA$ and
$AA$, and open up new prospects in TMD physics. 
In the same context, 
it is interesting to note that TMD physics has been demonstrated to be closely connected to the nucleon energy-correlator (NEEC)~\cite{Liu:2022wop} and the fragmenting energy correlator~\cite{Liu:2024kqt}, which can be expressed in terms of moments of TMD parton distributions~\cite{Liu:2024kqt} and has also been widely applied to studies for collider experiments recently~\cite{Cao:2023oef,Cao:2023qat,Li:2023gkh,Guo:2024jch,Guo:2024vpe,Chen:2024bpj}. Meanwhile, the same model mechanism should also apply to the NEEC, which could serve as another test of the model at currently available facilities such as HERA and CEBAF at JLab, or at future electron-ion colliders.

Looking ahead, a more sophisticated parameter fitting could improve the model's validation by exploring correlations between one-point energy flow and the jet axis. An identical Collins-Soper kernel is expected to appear in the energy-jet-axis correlator accompanied by an additional perturbative Sudakov factor. The abundant data from LHC and LEP experiments will facilitate a more precise determination of the kernel, further testing the model. Exploring the TMD parameterizations from other global analyses~\cite{Scimemi:2019cmh,Bacchetta:2022awv} would be desirable as well.

\begin{acknowledgments}


\textbf{\textit{Acknowledgements.}} 
We thank Chris Cocuzza, Andreas Metz, Volker Koch for discussions on subjects related to this work. 
We would like to thank the MITP for its hospitality when this work was initialized.  
This work is supported by the Office of Science of the U.S. Department of Energy under Contract No. DE-AC02-05CH11231. 
X.~L. is supported by the Natural Science Foundation of China under Contract No.~12175016. H.~X.~Z. is supported by the Natural Science Foundation of China under Contract No.~12425505.
This work has been supported in part by the Bundesministerium f\"{u}r Bildung und 
Forschung (BMBF) under grant no. 05P21VTCAA.

 \end{acknowledgments}

\bibliographystyle{h-physrev}   
\bibliography{refs}

\newpage
\newpage
\newpage
\appendix
\begin{widetext}
\section{Supplemental Material for ``Universality in the Near-Side Energy-Energy Correlator"}

\section{Non-perturbative features of the near-side EEC} 
We highlight the non-perturbative features of EEC in the small $\chi$ limit using Pythia simulation~\cite{Sjostrand:2014zea}. Some of the features have been also discussed in the experimental publications~\cite{CMS:2024mlf,ALICE:2024dfl}. We first present in Fig.~\ref{fg:full_pythia} the EEC spectrum measured globally in $e^+e^-$ annihilation events at various center of mass energies $Q$. This figure essentially plots the same EEC as in Fig.~\ref{fg:zoom-small-theta} but without rescaling the $x$-axis with $Q$. The Pythia results are normalized and compared to the OPAL data~\cite{OPAL:1993pnw} to validate the setup. 
\begin{figure}[htbp]
  \begin{center}
   \includegraphics[scale=0.4]{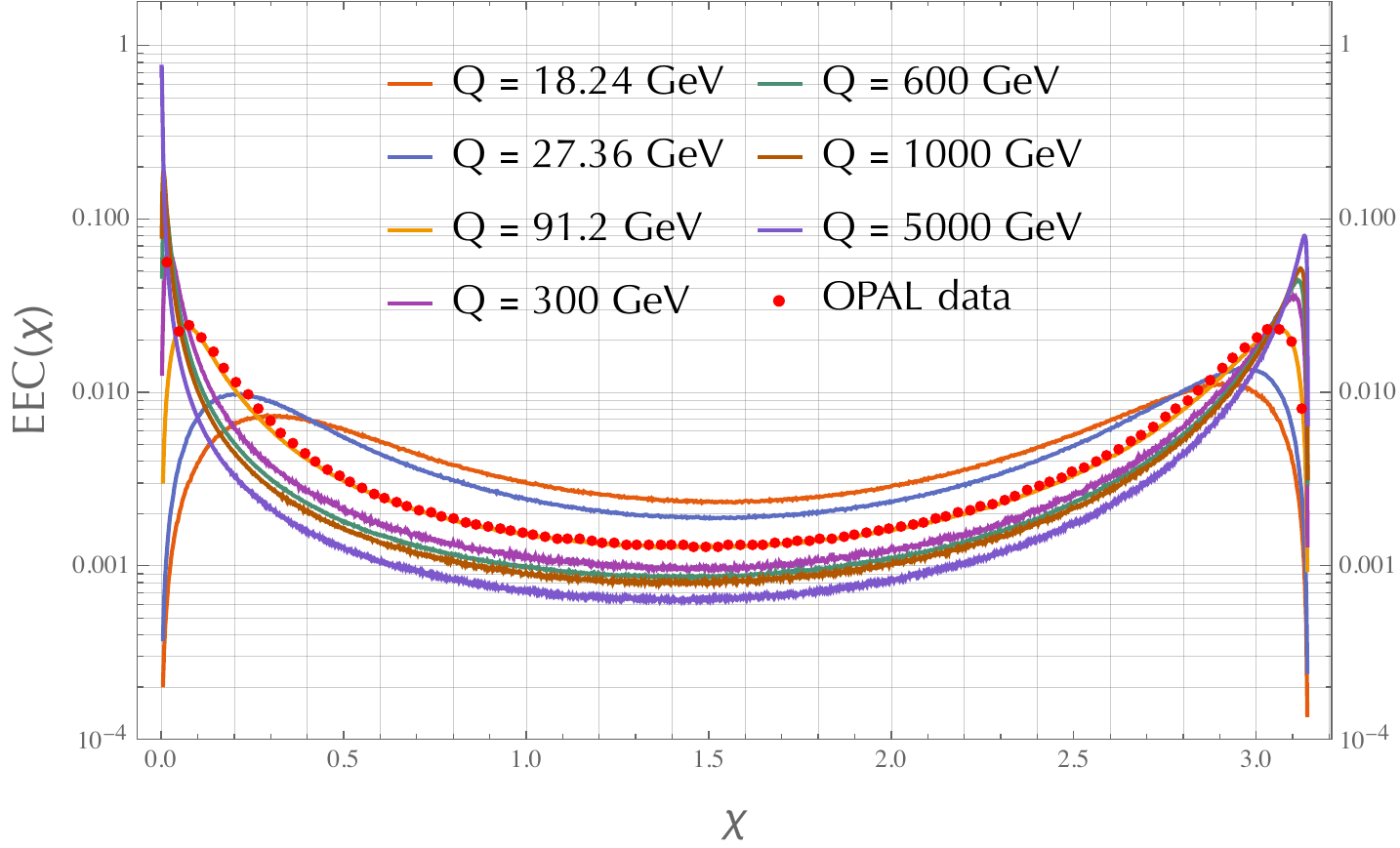} 
\caption{EEC spectrum in $e^+e^-$ annihilation for different values of $Q$. }
  \label{fg:full_pythia}
 \end{center}
  \vspace{-5.ex}
\end{figure}
A noteworthy feature we have emphasized in the main text is the differing behavior of the peaks in the regions $\chi \to 0$ and $\chi \to \pi$ when $Q$ varies. The former rises more sharply with increasing $Q$ compared to the ones as $\chi \to \pi$. The apparently symmetric shape observed at $Q = 91.2\>\rm{GeV}$~\cite{OPAL:1993pnw} is coincidental.

To verify the universality of the small $\chi$ behavior, in Fig.~\ref{fg:lhceec} we study the EEC measured within jets in $pp \to jj$. We set $\sqrt{s} = 14 \, {\rm TeV}$. 
We apply the anti-$k_t$ jet algorithm with a jet radius $R = 0.4$ to construct the jets and select quark jets. We measure the EEC using the leading jet in the central region $|\eta_J| < 2.1$ and require the jet energy $E_J \in (0.98 \frac{Q}{2}, 1.02 \frac{Q}{2})$, for various selected values of $Q$. The measured jet EEC has been normalized to the EEC in $e^+e^-$ annihilation. 
\begin{figure}[htbp]
  \begin{center}
   \includegraphics[scale=0.4]{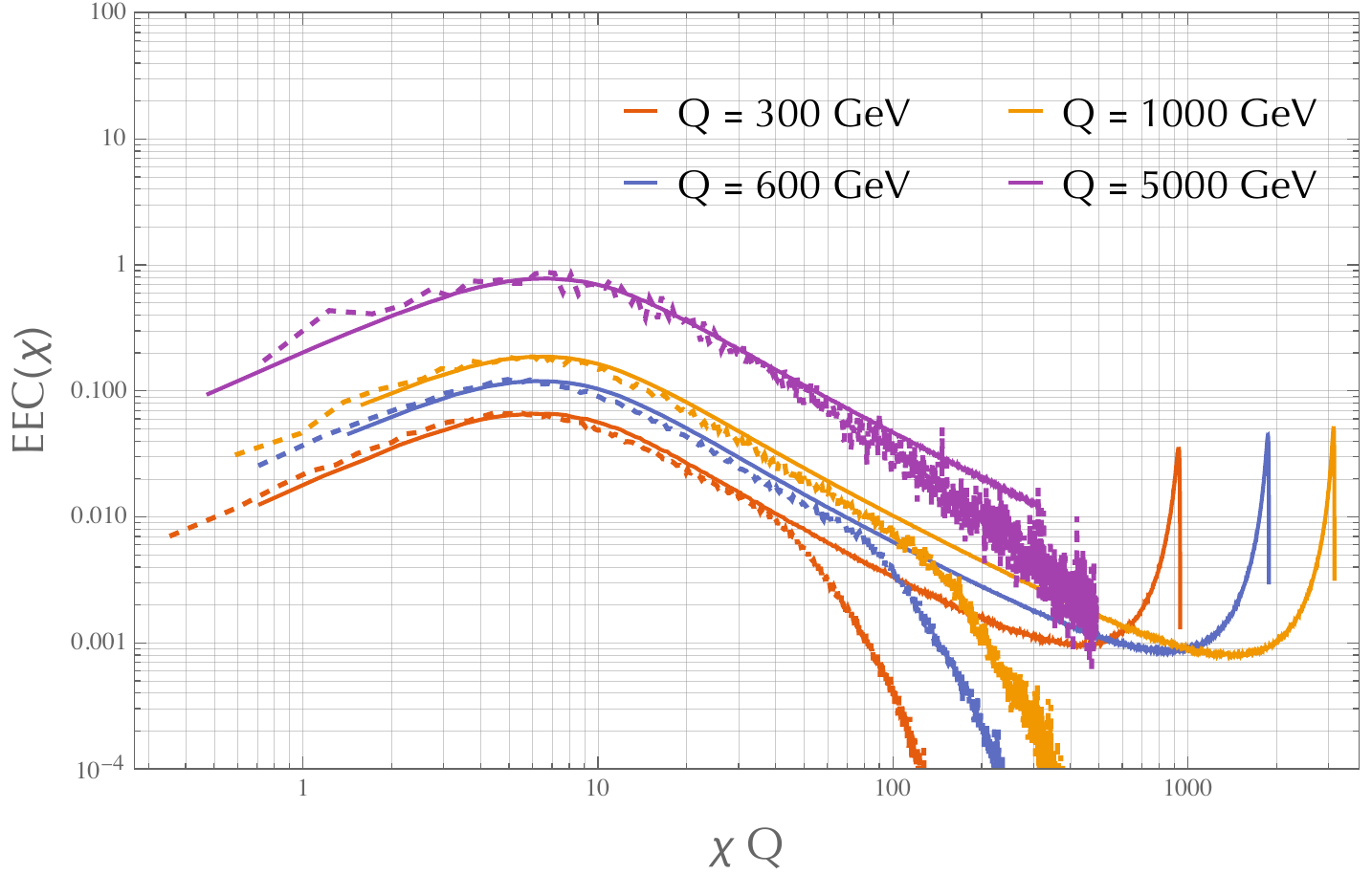} 
\caption{Comparison between the EEC in jets and measured globally. Solid lines represent the EEC in $e^+e^-$ annihilation and dashed curves the EEC  in quark jet in $pp$ collisions.}
  \label{fg:lhceec}
 \end{center}
  \vspace{-5.ex}
\end{figure}
Figure~\ref{fg:lhceec} shows good agreement between the EECs in jets and measured globally, suggesting the universality of the EEC at small $\chi$.
We note that for low values of $Q$, the jet EEC spectrum slightly shifts to the left compared to the global EEC. This is due to the choice of a small jet cone, even in the deep non-perturbative regime and, as we have checked, can be eliminated by choosing a larger jet radius. 

\section{Additional Comparisons between the Model, Pythia Simulation and Experimental Data} 

Here we present comparisons of our model to Pythia simulations for the EEC distribution measured in gluon jets. 
Reasonable agreement is obtained, as shown by Fig.~\ref{fg:model-g}. Especially the shapes and heights of the peaks 
in the non-perturbative region are well reproduced across all values of $Q$. Compared with the quark EEC, the peak position slightly shifts to larger values of $\chi Q$.   
\begin{figure}[htbp]
  \begin{center}
   \includegraphics[scale=0.4]{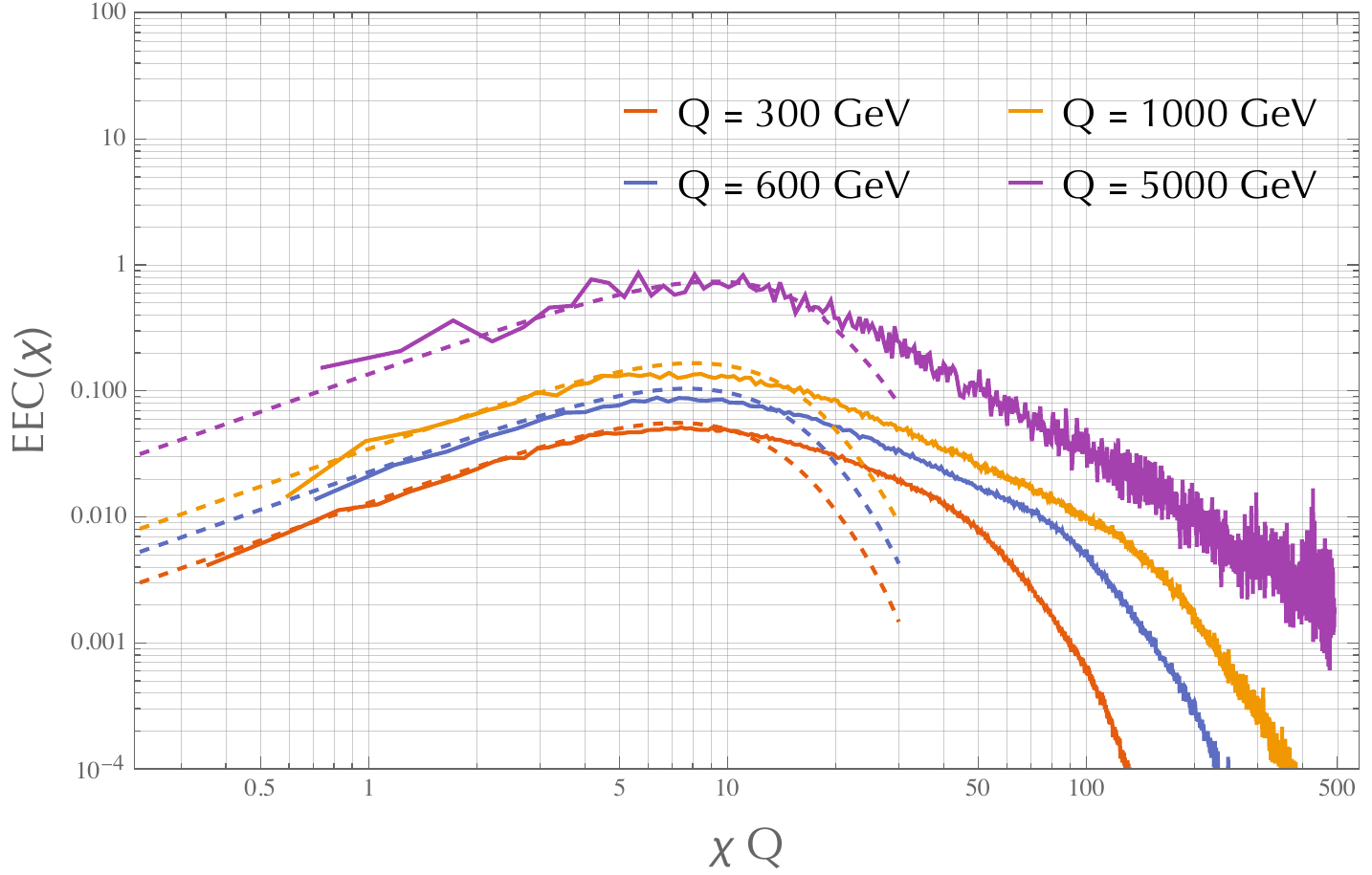} 
\caption{Comparison between the gluon EEC using Pythia simulation (solid curves) and model prediction (dashed curves). }
  \label{fg:model-g}
 \end{center}
  \vspace{-5.ex}
\end{figure}

Figure~\ref{fg:model-alice} shows a comparison between the model predictions and recent measurements by ALICE~\cite{ALICE:2024dfl}. Since ALICE normalized the EEC spectrum to the 
total number of jets
in the measured range, we have adjusted the model prediction to match the distribution in the lowest jet $p_T$ bin to determine the normalization factor, $N$.  
\begin{figure}[htbp]
  \begin{center}
   \includegraphics[scale=0.3]{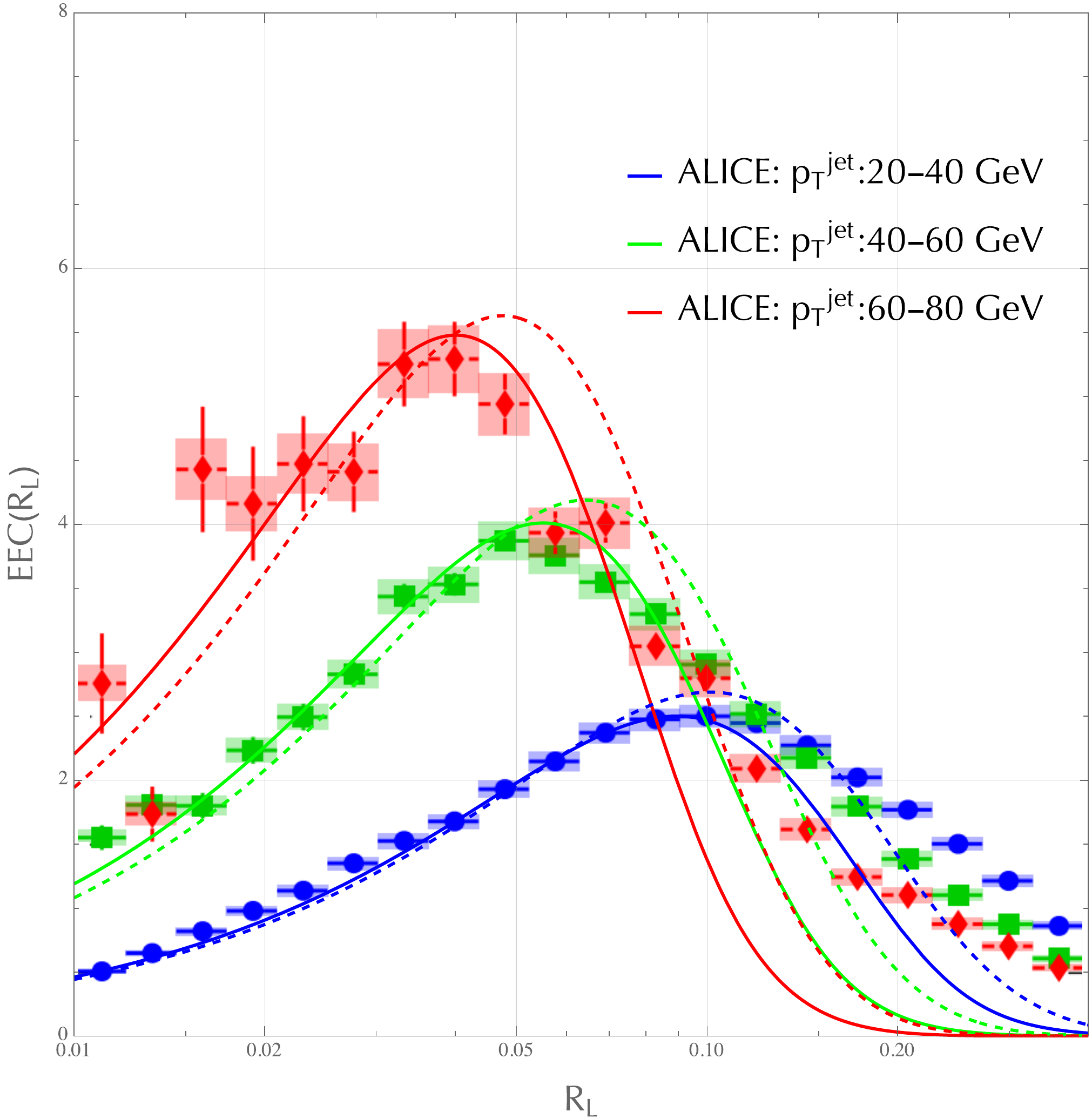} 
\caption{Comparison between the model predictions (curves) and the ALICE measurements extracted from Ref.~\cite{ALICE:2024dfl}. Here $R_L = \chi$. Solid and dashed lines refer to quark and gluon jets, respectively.}
  \label{fg:model-alice}
 \end{center}
  \vspace{-5.ex}
\end{figure}
It is important to note that ALICE employed $p_T$ weighting instead of energy weighting when constructing the EEC. 
We approximate the jet energy $E_J$ by the medium value of $p_T$ in each $p_{{\rm jet},T}$ bin. This substitution might underestimate the jet energy. Additionally, since the ALICE measurement used an inclusive charged jet sample, the exact quark/gluon jet fraction remains unclear to us at this stage, and in Fig.~\ref{fg:model-alice}, we present the EEC distribution for the quark jets (solid lines) and gluon jets (dashed curves) separately. 
From Fig.~\ref{fg:model-alice}, we observe a satisfactory agreement between the model prediction and the ALICE data, thereby further validating the proposed model.

Figure~\ref{fg:e3c_eec_cms} illustrates the ratio of E3C to EEC using Pythia simulation for both quark and gluon jets. The ratio in the perturbative region with $\chi Q > 5\, {\rm GeV}$ has been utilized to achieve the most precise $\alpha_s$ extraction using jet substructures~\cite{CMS:2024mlf}. In the non-perturbative region when $\chi \to 0$, the ratio tends to become independent of $\chi$ and the value is consistent with the model prediction $c\approx 0.7$, as manifest in Fig.~\ref{fg:e3c_eec_ratio}.
\begin{figure}[htbp]
\begin{subfigure}[h]{0.45\linewidth}
\includegraphics[width=\linewidth]{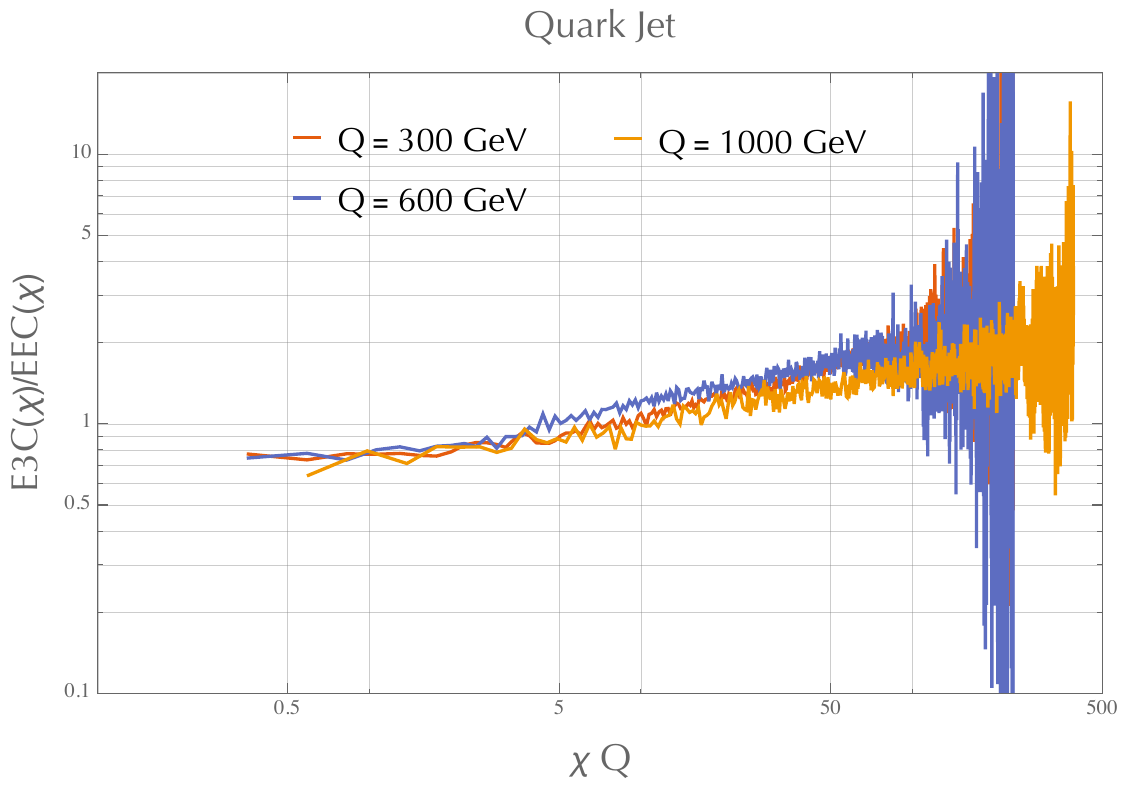}
\caption{E3C/EEC in quark jets}
\end{subfigure}
\hfill 
\begin{subfigure}[h]{0.45\linewidth}
\includegraphics[width=\linewidth]{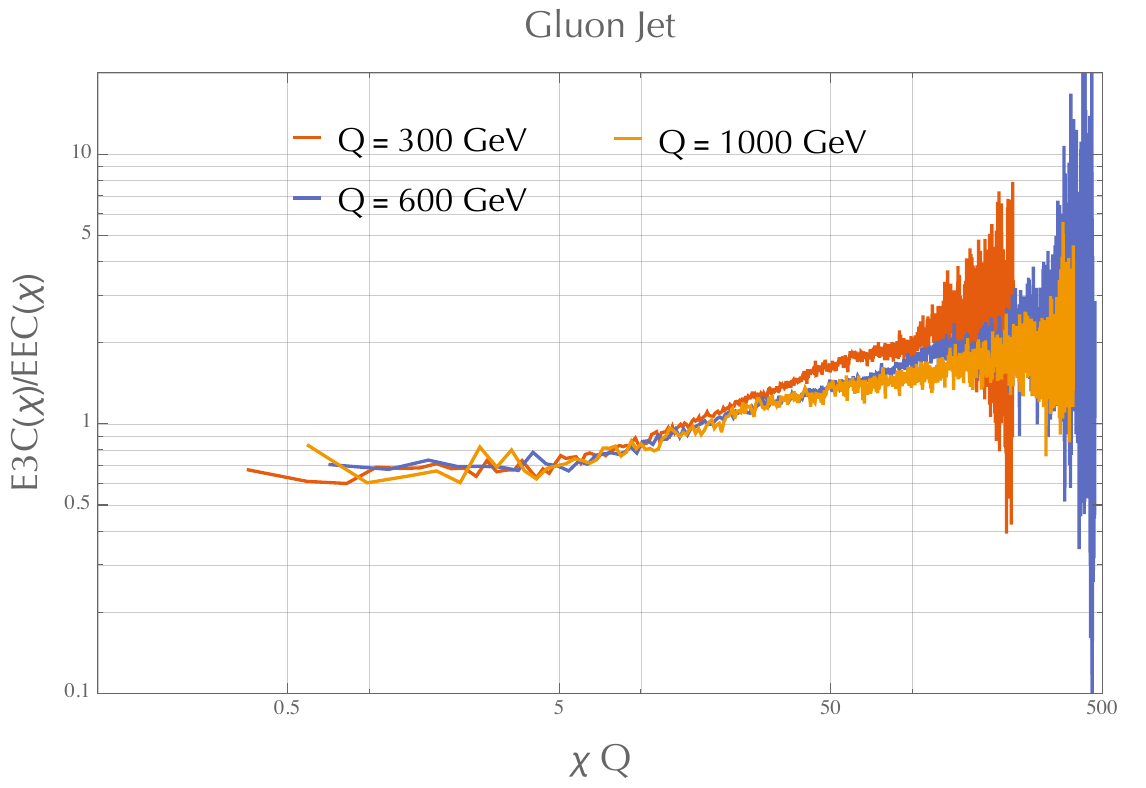}
\caption{E3C/EEC in gluon jets}
\end{subfigure}%
\caption{E3C/EEC from Pythia Simulation in quark and gluon jets.}\label{fg:e3c_eec_ratio}
\end{figure}

One can go one step further 
if one assumes that in the small $\chi$ limit, the 2 prongs with larger angular separation are uncorrelated and generate independent soft transverse kinematics of the third prong.  
Integrating over the third prong 
of the angular distribution, one finds
\bea\label{eq:model-ratio} 
\frac{\rm{E3C}(\chi)}{\rm{EEC}(\chi)} |_{\chi \to 0}
\approx 2c + c^3 E_J^2  \int \theta \, d\theta \, d\phi \, db b
\Theta(\chi-\theta) \Theta\big(\chi^2 - (\theta\cos\phi-\chi)^2-(\theta\sin\phi)^2\big)
J_0(c E_J \theta b)e^{-2 \frac{g_1}{c^2}b^2}
e^{-2 \frac{g_2}{2}\ln\left(\frac{b}{b_\ast} \right)\ln\frac{cE_J}{\mu_0}}\,. \nonumber \\ 
\eea 
Figure~\ref{fg:model-ratio} shows that this prediction is also in reasonable agreement with the CMS data away from $\chi=0$. 
\begin{figure}[htbp]
  \begin{center}
   \includegraphics[scale=0.34]{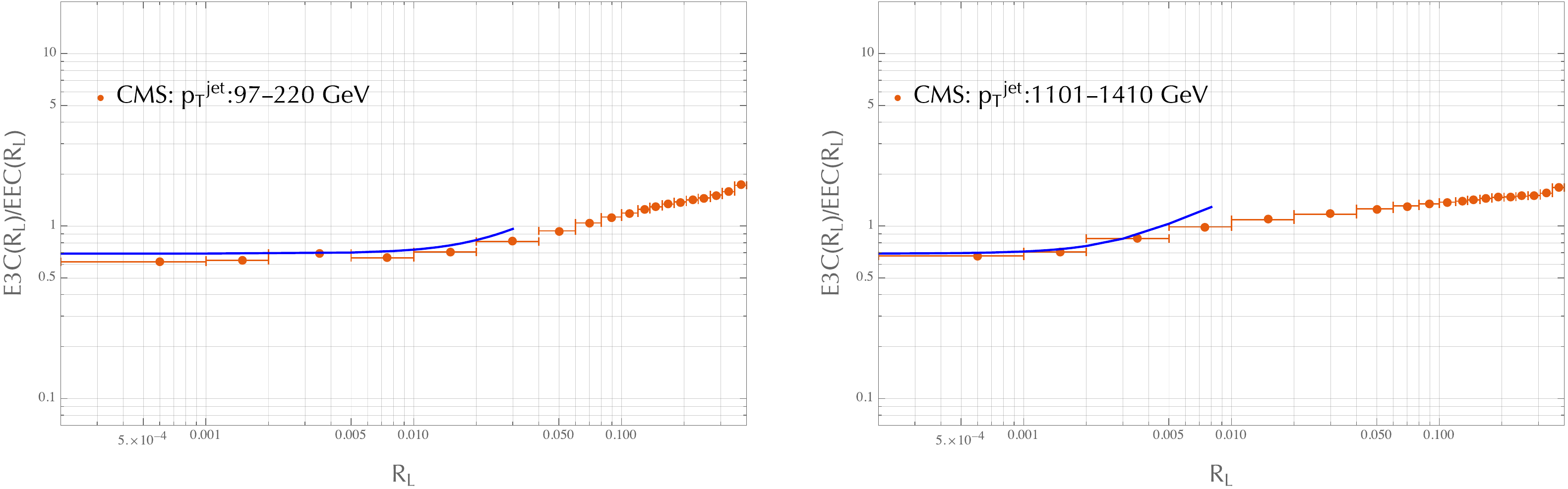} 
\caption{Comparison between the model prediction in Eq.~(\ref{eq:model-ratio}) (blue solid lines) and the CMS data~\cite{CMS:2024mlf}.}
  \label{fg:model-ratio}
 \end{center}
  \vspace{-5.ex}
\end{figure}

Last we present the prediction of the ratio in Figure~\ref{fg:model-ratio-alice} for the gluon EEC with ALICE kinematics.  
\begin{figure}[htbp]
  \begin{center}
   \includegraphics[scale=0.6]{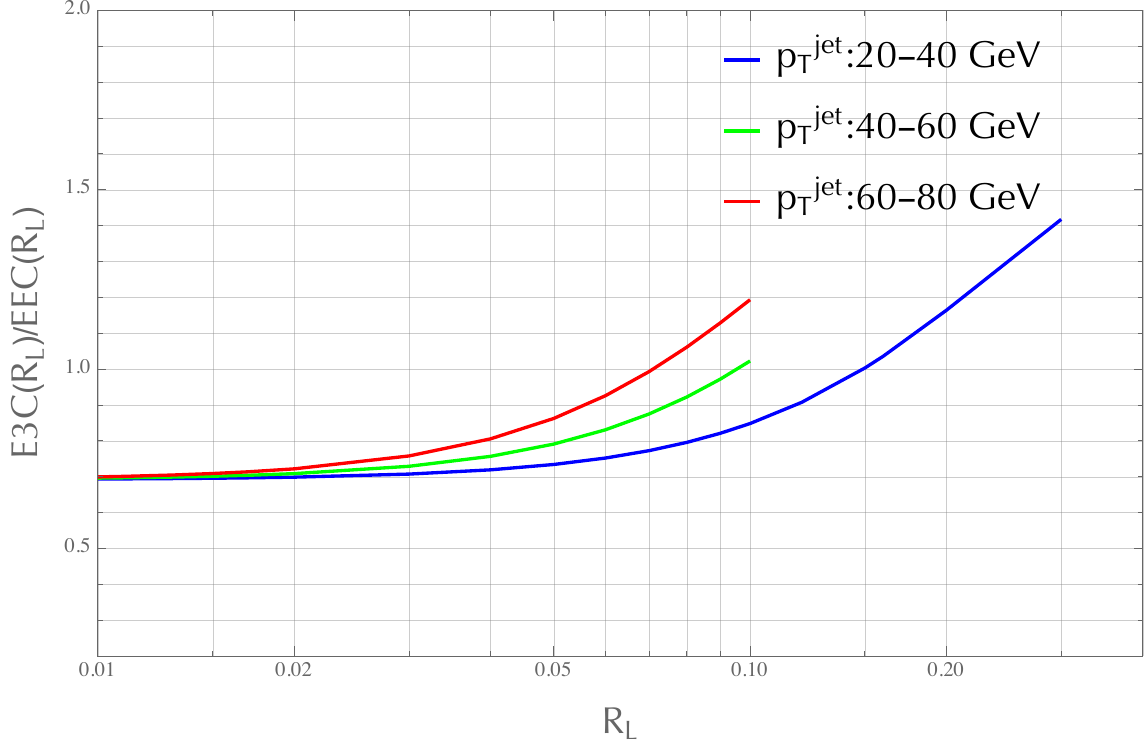} 
\caption{ Gluon E3C/EEC with ALICE settings. }
  \label{fg:model-ratio-alice}
 \end{center}
  \vspace{-5.ex}
\end{figure}

\end{widetext}

\end{document}